# Quantitative characterization of the imaging limits of diffuse low-grade oligodendrogliomas

C. Gerin[1,7], J. Pallud[2,5,8], C. Deroulers[1,6], P. Varlet[3,5], C. Oppenheim[4,5], F. X. Roux[2,5], F. Chrétien[3,5], S. R. Thomas[9], B. Grammaticos[1,7], M. Badoual[1,6]

1: IMNC Laboratory, UMR 8165, CNRS, Paris Sud and Paris Diderot Universities, Bât. 104, 91406 Orsay, France

2: Department of Neurosurgery, Sainte-Anne Hospital, Paris, France

3: Department of Neuropathology, Sainte-Anne Hospital, Paris, France

4: Department of Neuroradiology, Sainte-Anne Hospital, Paris, France

5: University Paris Descartes, Paris, France

6: University Paris Diderot, Paris, France

7: Centre National de la Recherche Scientifique (CNRS), France

8: Réseau d'Etude des gliomes, REG, Groland, France

9: IR4M Laboratory, UMR8081, CNRS, Paris Sud University, 91406 Orsay, France

Financial support: Comité des théoriciens de l'IN2P3, CNRS, France.

Corresponding author: Mathilde Badoual

Mailing address: Laboratoire IMNC, Campus universitaire d'Orsay, bat 440, 91405 Orsay, France

Phone/Fax: (33)169157201/(33)169157196

email: badoual@imnc.in2p3.fr

Disclosure of Potential Conflicts of Interest: No potential conflicts of interest were disclosed.

**Abstract.**

**Background** : Supratentorial diffuse low-grade gliomas in adults extend beyond maximal visible MRI-defined abnormalities, and a gap exists between the imaging signal changes and the actual tumor margins. Direct quantitative comparisons between imaging and histological analyses are lacking to date. However, they are of the utmost importance if one wishes to develop realistic models for diffuse glioma growth.

**Methods** : In this study, we quantitatively compare the cell concentration and the edema fraction from human histological biopsy samples (BSs) performed inside and outside imaging abnormalities during serial imaging-based stereotactic biopsy of diffuse low-grade gliomas.

**Results** : The cell concentration was significantly higher in BSs located inside (1189±378 cell/mm$^2$) than outside (740±124 cell/mm$^2$) MRI-defined abnormalities (p=0.0003). The edema fraction was significantly higher in BSs located inside (mean, 45±23%) than outside (mean, 5±9%) MRI-defined abnormalities (p<0.0001). At borders of the MRI-defined abnormalities, 20% of the tissue surface area was occupied by edema, and only 3% by tumor cells. The cycling cell concentration was significantly higher in BSs located inside (10±12 cell/mm$^2$) compared to outside (0.5±0.9 cell/mm$^2$) MRI-defined abnormalities (p=0.0001).

**Conclusions** : We show that the margins of T2-weighted signal changes are mainly correlated with the edema fraction. In 62.5% of patients, the cycling tumor cell fraction (defined as the ratio of the cycling tumor cell concentration to the total number of tumor cells) was higher at the limits of the MRI-defined abnormalities than closer to the center of the tumor. In the remaining patients, the cycling tumor cell fraction increased towards the center of the tumor.

## 1. Introduction

Diffuse gliomas are the most common primary brain tumors in adults. Diffuse low-grade gliomas (DLGG) (World Health Organization grade II gliomas [1]) share better outcomes than high-grade gliomas [2, 3, 4]. Despite refinements in imaging techniques and therapeutic management, DLGG remain incurable. Tumor invasiveness, particularly tumor cell proliferation and migration, thwarts the efficacy of oncological treatments. With magnetic resonance imaging (MRI), DLGG usually appear as non-contrast enhancing masses of hypointensity with T1- and of hyperintensity with T2-weighted and Fluid Attenuation Inversion Recovery (FLAIR) sequences [5, 6]. Since T2-weighted and FLAIR sequences have a high sensitivity to changes in water content of tissues, the hypersignal in such sequences is generally assumed to reflect the edema that surrounds tumor cells [7, 8, 9]. Conventional MRI underestimates the spatial extent of diffuse gliomas [10, 11, 12, 13], and we previously showed that conventional MRI underestimates the actual spatial extent of DLGG by identifying isolated tumor cells lying some centimeters beyond imaging abnormalities with T2-weighted and FLAIR sequences in DLGG [14]. The existence of such isolated tumor cells explains: 1) recurrences at the resection margins after gross total surgical removal [15]; 2) the prognostic significance of the extent of resection [16, 15, 17]; and 3) the fact that functional-based resections encompassing a margin beyond MRI-defined abnormalities improves the outcome of DLGG [18].

Modeling is an alternative way to estimate the actual spatial DLGG extent. A diffusion-proliferation partial differential equation has been proposed to model: 1) the evolution of the cell concentration in diffuse gliomas [19, 20]; 2) oncological treatment efficacy [21, 22, 23, 24, 25]; and 3) the date of birth of DLGG [26]. This model explains tumor recurrences as the consequence of the diffusion of tumor cells into the surrounding tissue, the density of these tumor cells being below the visibility threshold of MRI, including T2-weighted sequence. This model is based upon the assumption that MR signal changes are linked to tumor cell concentration, and thus MR signal changes should correspond to a cell concentration threshold [24]. This threshold has been measured once for a glioblastoma by CT-scan [22], but not by MRI and not for DLGG. A better understanding of the differences between MRI-defined abnormalities and the spatial extent of DLGG is mandatory in order to optimize therapeutic management and to develop models closer to the biological reality. The aim of the present study is to characterize quantitatively the histopathological abnormalities (cell concentration, edema) and to compare them with the imaging abnormalities. We analyzed biopsy samples (BSs) from serial MRI-based stereotactic biopsies performed within and beyond MRI-defined abnormalities in DLGG.

## 2. Material and methods

### 2.1. Data selection

We searched the database of a previously published study performed in our institution [14] and focused on patients for whom a low-grade oligodendroglioma was newly diagnosed by MRI-based serial stereotactic biopsies according to the Talairach stereotactic method between January 1992 and December 2001. Inclusion criteria were: 1) absence of oncological treatment prior to stereotactic biopsy; 2) supratentorial hemispheric location; 3) well delineated hypersignal on T2-weighted sequences; 4) absence of contrast enhancement [5]; 5) less than three-week interval between preoperative MRI and the stereotactic biopsy procedure; 6) biopsy samplings extending beyond MRI-defined abnormalities; and 7) available data for subsequent imaging and pathological analyses. These patients gave their informed consent for storage of the surgical samples for further analyses. We retrospectively selected nine cases of untreated adult patients (median age: 39 years, range 29-54 years) with a newly diagnosed DLGG by between June 1993 and September 2000. Clinical and imaging characteristics of tumors are presented in Table 1. We analyzed 44 BSs (median, 5; range, 3-8 per patient).

### 2.2. Histological and immunostaining techniques

The BSs were fixed in formalin-zinc (formalin 5%, zinc 3 g/L, sodium chloride 8 g/L) and individually embedded in paraffin. Serial sections were cut at 6 μm. Sections used for immunohistochemistry were microwaved in citrate buffer (pH=6) for antigen retrieval (Micro

MedMicro MEDT, Hacker instruments, Winnsboro, SC) at 98°C for 30 minutes. Ki-67 (MIB-1) immunostaining revealed MIB-1-positive cells (i.e., cycling cells) [27]. IDH1 immunolabeling was realized by a Ventana automatised system (Benchmark XT, Ventana Medical Systems, Tucson, USA). A standard pretreatment protocol included CC1 buffer and then an IDH1 R132H (1/50, clone H09, Dianova, Hamburg, Germany) incubation for 32 minutes at room temperature. Antibody binding was visualized with an Ultraview Universal Kit (Ventana, Tucson, USA). Diaminobenzidine tetra hydrochloride (DAB, Ventana) was used as the chromogen.

### 2.3. Digitization of the biopsy sample slices

All BSs slices were scanned and digitized with the same microscope (Hamamatsu NanoZoomer C9600-12), at several magnifications. The images have a resolution of 4.4 pixels/µm at the highest magnification (40x). The digitized images were subsequently analyzed using the free software ImageJ (version 1.43).

### 2.4. Fixing the limits of the MRI-defined abnormalities

The MRI-defined abnormalities were estimated using T2-weighted sequence. Possible sources of variability were the manual determination of the limits and the manual choice of the signal window, therefore, the definition of the MRI-defined abnormalities based only on the absolute gray levels would not be reproducible. To solve these problems, we used the following method: 1) we plotted the gray levels of the MRI along the biopsy trajectory; 2) we manually selected the part of the trajectory located inside the cortex, and we calculated the average gray level corresponding to the cortex signal (Figure 1); and 3) we defined the limit of the signal abnormality along the biopsy trajectory as the position where the gray levels drops below the average gray level of cortex.

To ensure the independence between the limits defined as explained above and the initial signal window, we applied this method on a series of images obtained through simulation with different signal windows (Figure 2). In this simulation, the MRI-defined abnormality with T2-weighted sequence of the tumor-brain system is assumed to be the sum of the contribution of the tumor, of the white matter, of the cortex, and of the cerebrospinal fluid. The whole system is discretized as a 100x100 matrix $0 \leq x \leq 100$, $0 \leq y \leq 100$. The tumor's contribution is modeled by a Gaussian function, $z(x,y) = 100 \exp(\frac{-(x-50)^2 + (y-50)^2}{100})$. The contribution of the white matter is modeled by a white noise of small amplitude, equal to z=10 (it can be seen in figure 2 as the dark gray area around the tumor). The cortex is modeled by a thin layer (ring-shaped) around the white matter, of constant amplitude z=20. The cerebrospinal fluid is modeled by a thin layer around the cortex, of constant amplitude z = 94.

We simulated three images from this model signal with different signal windows (contrasts): [0, 115], [0, 50], and [10, 30]. For instance, if the window is [0, 50], the cerebrospinal fluid (z=94) cannot be distinguished from the core of the tumor (z = 100), because both are mapped onto full white pixels (Figure 2, middle). In all three pictures, even though the contrast between white and gray matter changes greatly, the abnormality limit according to our procedure remained the same centered circle with diameter 26.5 ± 1 pixels.

For each patient, the MRI-defined abnormalities defined with this method were validated independently by a senior neurosurgeon (JP).

## 2.5. Imaging and pathological spatial correlation

For the purpose of the present study, we have reassessed the position of each BS with respect to MRI-defined abnormalities using images performed for clinical use. The position of a BS was obtained by reassessing its position in space using a superimposition of intraoperative teleradiographic x-ray images and preoperative reformatted MR images with T2-weighted sequences in the same plane as the biopsy trajectory. Postoperative MRI showing the biopsy trajectory was manually merged to verify the accuracy of image superimposition. The methodology is detailed in [14].

For each BS, the position in space with respect to the MRI-defined abnormalities with T2-weighted sequence was measured (Figure 1). A number was assigned to each BS, indicating its position relative to the limits of MRI-defined abnormalities along the biopsy trajectory: zero if the BS was on the limit, a positive value equal to the distance to the limit if the BS was outside MRI-defined abnormalities, and a negative value equal to the opposite of the distance to the limit if the BS was inside MRI-defined abnormalities (Figure 1). The uncertainty (error bar) on the position of each BS, which results from the uncertainty on the position of the whole biopsy trajectory, was estimated to be between 4 and 8 mm. This uncertainty results from the actual peroperative position of the whole biopsy trajectory, which could be 2 mm higher or lower than represented on the MRI slice.

## 2.6. Cell count

For each BS, cell count analyses were restricted to regions of infiltrated white matter (exclusion of red blood cells and endothelial cells). Only the green component of the picture, with the highest contrast between cells and tissue, was kept. The threshold was chosen so as to select only the cell nuclei. Each BS was divided into four comparable regions, and the cell concentration in each region was computed as the number of cells divided by the surface area. The mean cell concentration was calculated as the mean over the four regions. The error bars on the cell concentration represent the maximum and the minimum cell concentration among the four studied regions.

## 2.7. Edema quantification

To quantify the edema, we developed a new method based on the color analysis of pictures of digitized BSs.

Using the RGB model, a triplet of relative intensity levels of red, green, and blue (R,G,B) characterizes any color, each component intensity being coded on 256 intensity levels. The color "white" corresponds to the triplet (255, 255, 255) and the color "black" corresponds to (0,0,0).

Each pixel has its own color and its own triplet (R,G,B). The three-color histograms of pictures of digitized slices of BSs were obtained using the ImageJ plugin "Color Histogram". We considered that the slices were two-dimensional, since they were not thicker than a cell diameter. Images did not display a uniform color, so the color histograms, representing the distributions of levels of (R,G,B) over all the pixels of the picture, were broad (Figure 3). To characterize the tissue, we used the mean values of these distributions, and we referred to them as: $\overline{R}$ for red, $\overline{G}$ for green, and $\overline{B}$ for blue histograms. The color of the background was defined as ($\overline{R}_w, \overline{G}_w, \overline{B}_w$). Colored with hematoxilin and eosin, a normal tissue appeared mainly pink, corresponding to ($\overline{R}, \overline{G}, \overline{B}$) with $\overline{R} \approx \overline{B}$ and $\overline{G} < \overline{R}$. A tissue with edema appeared clearer than the normal brain. We defined the edema fraction $x$ as the ratio of the surface area occupied by edema to the surface area of the BS. The tissue with edema corresponded to the levels :

$(\overline{R}_e, \overline{G}_e, \overline{B}_e) = (\overline{R}(1 - x) + \overline{R}_w x, \overline{G}(1 - x) + \overline{G}_w x, \overline{B}(1 - x) + \overline{B}_w x)$

Therefore,

$x = ((\overline{R} - \overline{G}) - (\overline{R}_e - \overline{G}_e))/((\overline{R} - \overline{G}) - (\overline{R}_w - \overline{G}_w))$

The quantities $\overline{R}_e$, $\overline{G}_e$ and ($\overline{R}_e - \overline{G}_e$) varied linearly with $x$. We decided to use the quantity $\overline{R}_e - \overline{G}_e$ to quantify edema, because the difference between two color components was less sensitive to differences between patients. The quantity $\overline{R} - \overline{G}$ for a normal tissue is close to 100 and $\overline{R}_w - \overline{G}_w$ for the background is close to 0, so the calibration function, for a given patient is close to the function: $x = 1 - 10^{-2}(\overline{R}_e - \overline{G}_e)$. Since pathological samples were prepared using the same routines, we assumed that all the BSs for one patient were colored the same way, so that the difference of coloration between the BSs were only due to different amounts of edema. Each BS was divided into four regions, and the color histograms of each region were extracted. The most distant BS from MRI-defined abnormalities was considered as "normal" by the pathologists, with an edema fraction equal to zero (sample P1 in Figure 1), and the means of the green and the red histograms were used to calculate R−G. For the other BSs *Pi* with *i > 1*, the edema fraction was obtained with the formula:

$$x(i) = ((\overline{R} - \overline{G}) - (\overline{R_e} - \overline{G_e})) / ((\overline{R} - \overline{G}) - (\overline{R_w} - \overline{G_w}))$$

In order to estimate the error due to the variation of staining from sample to sample, we prepared new slices from the BSs of a patient (patient 6). We found that the variation of the estimated fraction of edema does not exceed 6% between the old slices and the new ones (supplementary figure).

As explained above, the error bars on the edema fraction represented the maximum and the minimum edema fraction among the four studied regions (intra-sample variability). However, to also take into account the sample-to-sample variability estimated around 6%, we fixed the minimum length of the error bars to +/- 6%.

## 2.8. Statistical analysis

Descriptive results are presented as mean ± standard deviation (range). Analyses were carried out using the unpaired t-test or Mann-Whitney rank-sum test for continuous variables, as appropriate. A P-value of less than 0.05 was considered to be significant. All statistical analyses were performed using JMP, version 10.0.0 (SAS institute, Cary, NC, USA). For linear regressions, the coefficient of determination was calculated.

## 3. Results

### 3.1. Edema fraction

For all patients, the mean edema fraction of each BS was plotted as a function of the position of the BS along the axis of the biopsy trajectory (Figure 4). The edema fraction was significantly higher in BSs located inside (mean 45 ± 23 %, range 11.1-88.0 %) compared to outside (mean 5 ± 9 %, range 0-33 %) MRI-defined abnormalities (p<0.0001). For all patients, the edema fraction was equal to zero in the BS situated at the greatest distance from the MRI-defined abnormalities. Although maximum edema fractions inside imaging abnormalities varied from one patient to one another, the edema fraction at the limit of the signal abnormalities on T2-weighted sequence (obtained by linear interpolation between two data points inside and outside MRI-defined abnormalities) corresponded to a common value of 20%.

### 3.2. Cell concentration

For all patients, the mean cell concentration of each BS was plotted as a function of the position of the BS along the axis of the biopsy trajectory (Figure 5).

In all but two patients (patients 1 and 6), the cell concentration increased from outside to inside MRI-defined abnormalities with a significantly higher cell concentration in BSs located inside (1189 ± 378; range 447-1900 cell/mm$^2$) compared to outside (740 ± 124; range 532-938 cell/ mm$^2$) MRI-defined abnormalities (p=0.0003). The mean value of the cell concentration at the limit of MRI-defined abnormalities (obtained by linear interpolation between two data points inside and outside MRI-defined abnormalities) was 981 ± 218 cell/mm$^2$ (range 750-1350 cell/mm$^2$). We assumed the "normal" cell concentration in the white matter (756 ± 111; range 577-893 cell/mm$^2$) as the cell concentration at the greatest distance from MRI-defined abnormalities. We assumed that the increase of cell concentration was due only to tumor cells, and we deduced the mean tumor cell concentration by subtracting the mean "normal" cell concentration. We obtained a mean tumor cell concentration at the limit of MRI-defined abnormalities of 225 ± 244 cell/mm$^2$ (range 0-600

cell/mm$^2$). The imaging signal changes did not seem to correspond to a unique cell concentration threshold (and unlikely to a tumor cell concentration threshold) among all patients, because the dispersion of the cell concentration at the limit of MRI-defined abnormalities was large.

We performed additional IDH1 immunostainings to identify mutated tumor cells in 3 BSs of patient 2 [28]. Since the cytoplasm of the cells was stained, we were able to estimate the tumor cell section area. We measured a maximum tumor cell section area of 206 μm$^2$ (n=188 cells (range 30-206 μm$^2$). Since the distribution of cell radii on digitized slides was uniform, we could estimate that the mean tumor cell section area was equal to two thirds of the maximum section area, i.e., 137 μm$^2$. At the limit of the MRI-defines abnormalities, a cell concentration of 225 cell/mm$^2$ (range 0-600 cell/mm$^2$) thus occupies a mean surface area of 3% (range 0-8%), seven times smaller than the surface area occupied by edema (around 20%). Therefore, edema contributed much more than tumor cells to the signal abnormalities and, hence, to the position of the abnormality limits with T2-weighted sequence.

We plotted the cell concentration related to the surface area of tissue (or corrected cell concentration) (Figure 5, top right). The "surface area of tissue" was defined as the difference between the whole surface area of the BS and the surface area of edema. In all but one patients (patient 1), the corrected cell concentration increased from outside to inside MRI-defined abnormalities with a significantly higher corrected cell concentration in BSs located inside (3097 ± 2975; range, 18-4876 cell/ mm$^2$) compared to outside (781 ± 130; range, 16-850 cell/ mm$^2$) MRI-defined abnormalities ($p<0.0001$). For most patients, edema appeared as large vacuoles devoid of any cell. In the tissue between vacuoles, the cell concentration can be very high. We directly measured, on the BS located at x = -15 mm of patient 4, that the local cell concentration is 104 cell/mm$^2$ in small tissue areas between edema vacuoles, in good agreement with the cell concentration calculated in Figure 5, top right.

### 3.3. Cycling cell concentration

For all patients, the mean cycling cell concentration of each BS was plotted as a function of the position of the BS along the axis of the biopsy trajectory (Figure 6, top left). The cycling cell concentration increased from outside to inside MRI-defined abnormalities, with a significantly higher cycling cell concentration in BSs located inside (10 ± 12; range, 0-41.3 cell/mm$^2$) compared to outside (0.5 ± 0.9; range, 0-3.9 cell/mm$^2$) MRI-defined abnormalities (p=0.0001). The corrected cycling cell concentration (where the surface area of edema was deduced from the total surface area of the tissue) also increased from outside to inside MRI-defined abnormalities with a significantly higher corrected cycling cell concentration in BSs located inside (32 ± 58; range, 0-230 cell/mm$^2$) compared to outside (0.6 ± 1.0; range, 0- 3.9 cell/mm$^2$) MRI-defined abnormalities (p<0.0001) (Figure 6, top right). We assumed that the increase in the cycling cells concentration was due only to the presence of tumor cells. We justify this assumption by the fact that it has been shown in [14] that MIB1-positive cells in these patients coexpressed OLIG2, and thus belong to the oligodendroglial lineage (which exclude reactive astrocytes or activated microglia). In the normal adult brain there exists a population of cycling cells (belonging to the oligodendroglial lineage) scattered throughout the cortex and the white matter [29], but their concentration is so low (the maximum is 0.4 cell/mm$^2$, or 0.05% of the total cell number) than it can be neglected compared to the cycling cell concentrations measured in the tumors.

Consequently, cycling tumor cells infiltrating the parenchyma around the tumor core were found up to 20 mm outside the MRI-defined abnormalities in accordance with previous reports [11, 13, 14]. We estimated the total tumor cell concentration (cycling and non-cycling tumor cells) by subtracting the normal cell concentration (measured in regions far from MRI-defined abnormalities) from the total cell concentration (corrected by edema) while assuming that normal cells did not die and did not proliferate inside the tumor core. We were able to calculate the cycling tumor cell concentration for Patient 1 at only one location (-16.4mm) since the sample at -5.7mm did not display any significant increase in the cell concentration compared to the normal tissue. Since one point is not sufficient to see a trend, we excluded this patient (but only from the cycling tumor cell study).

The cycling tumor cell fraction (defined as the ratio of the cycling tumor cell concentration to the total number of tumor cells) was significantly higher in BSs located inside (2 ± 2.6 %; range, 0-10 %) than outside (0.2 ± 0.3 %, range, 0-0.8 %) MRI-defined abnormalities (p=0.015) (Figure 6, bottom).

The eight remaining patients were divided into two groups: in 62.5% of cases (patients 2, 3, 4, 7, 8), the cycling tumor cell fraction was higher at the limit of the MRI-defined abnormalities compared to close to the center of the tumor. Therefore, cycling cells formed a ring of proliferation in the peripheral areas of the tumor. In 37.5% of cases (patients 5, 6, 9), the cycling tumor cell fraction increased towards the center of the tumor core. Tumors with a maximum cycling cell concentration at the limits of the MRI-defined abnormalities were larger (mean diameter 4.7 cm, ± 0.5 cm; range 4.0-5.3 cm) than those with a maximum cycling cell concentration at the center of the tumor (mean diameter 3.9 cm, ± 1.3 mm; range 2.8-5.3 cm) without reaching significance (p=0.365).

We also found a better correlation between the edema fraction (for a fraction of edema smaller than 80%) and the corrected cell concentration ($R^2 = 0.78$) than between edema and cell concentration (determination coefficient $R^2 = 0.32$), edema and cycling cell concentration (determination coefficient $R^2 = 0.03$), or edema and corrected cycling cell concentration (determination coefficient $R^2 = 0.05$). Since edema was linked to the corrected tumor cell concentration, we suggest that the presence of edema was related to the presence of tumor cells.

## 4. Discussion

In the present work, we quantified the edema fraction, the cell concentration and the cycling cell concentration both inside and outside MRI-defined abnormalities with T2-weighted sequence in adults that underwent MRI-based serial stereotactic biopsies for the diagnosis of untreated supratentorial diffuse low-grade oligodendrogliomas. We found that: 1) MRI-defined abnormalities with T2-weighted sequence were mainly correlated with the edema fraction but not with the total cell concentration, the tumor cell concentration, or to the cycling cell concentration; 2) the limits of the imaging abnormalities corresponded to a common threshold of 20% of edema fraction; and 3) the cycling cells formed a ring of proliferation at the DLGG borders, i.e., at the inner limits of the imaging abnormalities. In addition, we provided a new method to quantify edema on histological samples.

In the case of DLGG, we propose that edema results from alterations of the extracellular matrix due to the presence of infiltrative tumor cells, since we showed that the edema fraction was correlated to the corrected tumor cell concentration. Vacuoles of various sizes form between fibers [30, 31]. Since the blood-brain barrier is morphologically and radiologically intact in DLGG, the interstitial edema is more likely due to an alteration of the osmotic gradient between plasma and interstitial fluid [7]. Nevertheless, the mechanisms underlying the DLGG-related edema are still poorly understood. DLGG-related edema could be multifactorial and various factors have been demonstrated to modulate blood-brain barrier permeability. Invasive tumor cells can change the micro-environment by: 1) producing extracellular matrix polysaccharides; 2) secreting cytokines and growth factors [32]; 3) degrading the extracellular matrix and basal lamina by matrix metalloproteinases…; or 4) inducing aquaporin 4 expression leading to modification of water and electrolytes transcapillar exchanges and of ion homeostasis ([33], [34], [35]). All these modifications are thus more related to a cytotoxic than vasogenic general mechanisms. However, a recent study, based on a recent real-time in vivo imaging technique, showed that single isolated glioma cells were able to induce vascular remodeling and intussusceptive microvascular growth [36]. This study, revealing discrete peritumoral angiogenesis, suggests that in "pre-contrast enhancing" regions, tiny vascular changes could explain glioma-related brain edema.

Our results are in agreement with a previous report showing that the hyperintense region with T2-weighted sequence contained more water than normal tissue [31]. This study also reports a close visual correlation between peritumoral hyperintensity with T2-weighted sequence and areas of pallor on the hematoxylin-eosin stained whole-brain sections, but no quantitative measurements were performed.

We developed an original method of edema quantification. Its main advantage is that it does not involve any threshold, thus improving reproducibility. The main drawback is the assumption that the observed color changes are only due to the presence of edema, since the slice thickness and the coloration techniques were constant in all the BSs of a given patient.

The cycling tumor cell fraction varies within the tumor, with a maximal cycling tumor cell concentration at the inner limits of the MRI-defined abnormalities in 62.5% of cases and at the center of the tumor in 37.5% of cases. As a practical consequence, pathologists should pay particular attention to both the center and the peripheral regions of a DLGG to accurately estimate proliferation rates. In addition, since DLGG are infiltrative tumors made of isolated tumor cells that permeate the brain parenchyma, the proliferation rates should ideally be estimated by the ratio between cycling cells and tumor cells and not between cycling cells and the whole cell population, because the presence of numerous non-tumor cells may lead to underestimate the actual proliferation rates.

The finding that cycling cells form a ring of proliferation at the tumor peripheral areas recalls a previous study in which a maximum of metabolism was detected in BSs performed in the periphery of DLGG as compared to those performed in the tumor center [37]. The diffusion-proliferation model for diffuse gliomas [19, 22, 23] has also predicted a saturation phenomenon in the interior of some tumors: the cell proliferation decreases in regions with high cell concentration and is absent when the cell concentration is maximal. The evidence that such a saturation phenomenon occurs for some patients would mean that the cell concentration, even in some DLGG, is high enough to slow down the proliferation rate of tumor cells. In DLGG, this saturation phenomenon was unexpected, since these tumors do not display any neoangiogenesis. The hypothesis of saturation is reinforced by the fact that tumors with a maximum cycling cell concentration within the inner limits of the MRI-defined abnormalities tend to be larger than those with a maximum cycling cell concentration inside the tumor core. The high cell concentration suggests that tumor cells may start having a limited access to nutrients and oxygen but may remain under the hypoxic threshold triggering neoangiogenesis. Finding high cell concentrations in the interior of DLGG may help to explain the observed transformations towards higher grades of malignancy.

These results should help to conceive new models based on the findings that: 1) edema is more related to the imaging abnormalities than to cell concentration; 2) edema is correlated with tumor cell concentration. In addition, they may help to validate existing models if observed and predicted cell and cycling cell concentrations are compared.

**Acknowledgments**

We thank Philippe Bertheau, David Ameisen and Fatiha Bouhidel for the opportunity to use their slide scanner Nanozoomer and their kind help with it, Christophe Nioche, Emmanuel Mandonnet and Catherine Daumas-Duport for helpful discussions. We also thank Jean-François Meder, head of the neuroradiological department of the Sainte-Anne Hospital.

The group belongs to the CellTiss consortium.

**References**

[1] Louis David N., Ohgaki Hiroko, Wiestler Otmar D., et al. The 2007 WHO Classification of Tumours of the Central Nervous System. *Acta Neuropathologica* 2007;114:97–109.

[2] Soffietti R., Baumert B. G., Bello L., et al. Towards personalized therapy for patients with glioblastoma. *Eur. J. Neurol.* 2010;17:1124-1133.

[3] Ricard D., Idbaih A., Ducray F., Lahutte M., Hoang-Xuan K., Delattre J. Y. Primary brain

tumours in adults. *Lancet* 2012;26:1984-1996.

[4] Jaeckle K. A., Decker P. A., V. Ballman K., et al. Transformation of low grade glioma and correlation with outcome: an NCCTG database analysis. *J. Neurooncol.* 2004;104:253-259.

[5] Pallud J., Capelle L., Taillandier L., et al. Prognostic significance of imaging contrast enhancement for WHO grade II gliomas. *Neuro-oncology* 2009;11:176.

[6] Connor S. E., Gunny R., Hampton T., O'gorman R. Magnetic resonance image registration and subtraction in the assessment of minor changes in low grade glioma volume. *Eur. Radiol.* 2004;14:2061-2066.

[7] Kaal C. A., Vecht C. J. The management of brain edema in brain tumors. *Curr. Opin. Oncol.* 2004;16:563-600.

[8] Cha S. Neuroimaging in neuro-oncology. *Neurotherapeutics* 2009;6:465-477.

[9] Price S. J. Advances in imaging low-grade gliomas. *Adv. Tech. Stand. Neurosurg.* 2010;35:1-34.

[10] Lunsford L. D., Martinez A. J., Latchaw R. E. Magnetic resonance imaging does not define tumor boundaries. *Acta Radiol. Suppl.* 1986;369:154-156.

[11] Kelly P. J., Daumas-Duport C., Kispert D. B., Kall B. A., Scheithauer W., Illig J. Imaging-based stereotaxic serial biopsies in untreated intracranial glial neaplasms. *J Neurosurg.* 1987;66:865- 874.

[12] Iwama T., Yamada H., Sakai N., et al. Correlation between magnetic resonance imaging and histopathology of intracranial glioma. *Neurological Research* 1991;13:48-54.

[13] Watanabe M. Takeda N. Magnetic resonance imaging and histopathology of cerebral gliomas. *Neuroradiology* 1992;34:463–469.

[14] Pallud J., Varlet P., Devaux B., et al. Diffuse low-grade oligodendrogliomas extend beyond MRI-defined abnormalities. *Neurology* 2010;74:1724-1731.

[15] Pallud J. Diffuse Low-Grade Gliomas: What Does "Complete Resection" Mean? Tumors of the Central Nervous System, Volume 2. 2011:1rst ed., Springer, 153-163.

[16] Sanai N., Berger M. S. Glioma extent of resection and its impact on patient outcome. *Neurosurgery* 2008;4:753-764.

[17] Han S.J., Sughrue M. E. The rise and fall of "biopsy and radiate": a history of surgical nihilism in glioma treatment. *Neurosurg. Clin. N. Am*. 2012;23:207-214.

[18] Yordanova Y. N., Moritz-Gasser S., Duffau H. Awake surgery for WHO Grade II gliomas within "noneloquent" areas in the left dominant hemisphere: toward a "supratotal" resection. *J. Neurosurg*. 2011;115:232-239.

[19] Murray J. D. Mathematical biology. II: Spatial models and biomedical applications. Berlin: Springer-Verlag3rd ed. 2002.

[20] Cruywagen G.C., Woodward D.E., Tracqui P., Bartoo Grace T., Murray J. D., Alvord E. C. The modelling of diffusive tumours. *J. Biological Systems* 1995;3:937–945.

[21] Cook J., Woodward D.E., Tracqui P., et al. Resection of gliomas and life expectancy. *J. Neurooncol*. 1995;24:131.

[22] Tracqui P., Cruywagen G. C., Woodward D. E., Bartoo G. T., Murray J. D., Alvord E. C. A mathematical model of glioma growth: the effect of chemotherapy on spatio-temporal growth. *Cell Prolif.* 1995;28:17–31.

[23] Swanson K. R., Alvord E. C., Murray J. D. A quantitative model for differential motility of gliomas in gray and white matter. *Cell Prolif.* 2000;33:317-329.

[24] Harpold H. L., Jr E. C. Alvord, Swanson K. R. The evolution of mathematical modeling of glioma proliferation and invasion. *J. Neuropathol. Exp. Neurol*. 2007;1:1-9.

[25] Badoual M., Deroulers C., Aubert M., Grammaticos B. Modelling intercellular communication and its effect on tumour invasion. *Phys. Biol*. 2010;7:046013.

[26] Gerin C., Pallud J., Grammaticos B., et al. Improving the time-machine: estimating date of birth of grade II gliomas. *Cell. Prolif.* 2012;45:76–90.

[27] Scholzen T, Gerdes J. The Ki-67 protein: from the known and the unknown. *J. Cell. Physiol*. 2000;182:311-322.

[28] Ichimura K., Pearson D.M., Kocialkowski S., et al. IDH1 mutations are present in the majority of common adult gliomas but rare in primary glioblastomas. *Neuro. Oncol.* 2009;11:341-347.

[29] Geha S., Pallud J., Junier M.P., Devaux B., Leonard N., Chassoux F., et al. NG2+/Olig2+ cells are the major cycle-related cell population of the adult human normal brain. Brain Pathol. 2010;20:399-411.

[30] Aleu F., Samuels S., Ransohoff J. The pathology of cerebral edema associated with gliomas in man. Report based on ten biopsies. *Am J Pathol*. 1966;48:1043-1061.

[31] Tovi M, Hartman M, Lilja A, Ericsson A. MR imaging in cerebral gliomas. Tissue component analysis in correlation with histopathology of whole-brain specimens. *Acta Radiol.* 1994;35:495- 505.

[32] Hatashita S., Hoff J. T., Salamat S. M. An osmotic gradient in ischemic brain edema. *Adv. Neurol*.1990;52:85–92.

[33] Kelly P. J. Gliomas: Survival, origin and early detection. *Surg. Neurol. Int.* 1987;96:1-7.

[34] Habela C.W., Ernest N.J., Swindall A.F., Sontheimer H. Chloride accumulation drives volume dynamics underlying cell proliferation and migration. *J. Neurophysiol*. 2009;101:750-757.

[35] Warth A., Mittelbronn M., Wolburg H. Redistribution of the water channel protein aquaporin-4 and the K+ channel protein Kir4.1 differs in low- and high-grade human brain tumors. *Acta Neuropathol.* 2005;109:418-26.

[36] Winkler F., Kienast Y., Fuhrmann M ., et al. Imaging Glioma Cell Invasion In Vivo Reveals Mechanisms of Dissemination and Peritumoral Angiogenesis. *Glia* 2009;57:1306-15.

[37] Lamari F., La Schiazza R., Guillevin R., et al. Biochemical exploration of energetic metabolism and oxidative stress in low grade gliomas: central and peripheral tumor tissue analysis. [Article in French] *Ann. Biol. Clin.* (Paris). 2008;66:143-150.

**Figures and Tables:**

| Patient number | 1 | 2 | 3 | 4 | 5 | 6 | 7 | 8 | 9 |
|---|---|---|---|---|---|---|---|---|---|
| Age | 43 | 54 | 29 | 41 | 47 | 34 | 39 | 30 | 30 |
| Sex | M | F | M | F | F | M | M | F | M |
| Tumor location | Frontal | Fronto-temporal-insular | Fronto-parietal | Temporo-insular | Parietal | Frontal | Insular | Parietal | Frontal |
| Number of BSs | 4 | 4 | 3 | 5 | 3 | 4 | 4 | 4 | 3 |
| VDE (mm/yr) | 3.6 | 1.3 | 3.2 | - | 1.1 | - | - | 3.6 | - |
| Diameter (mm) | 6.0 | 4.6 | 4.0 | 4.8 | 2.8 | 5.3 | 4.8 | 5.3 | 3.7 |

**Table 1**: Clinical and imaging characteristics at histological diagnosis. VDE is the velocity of diametric expansion.

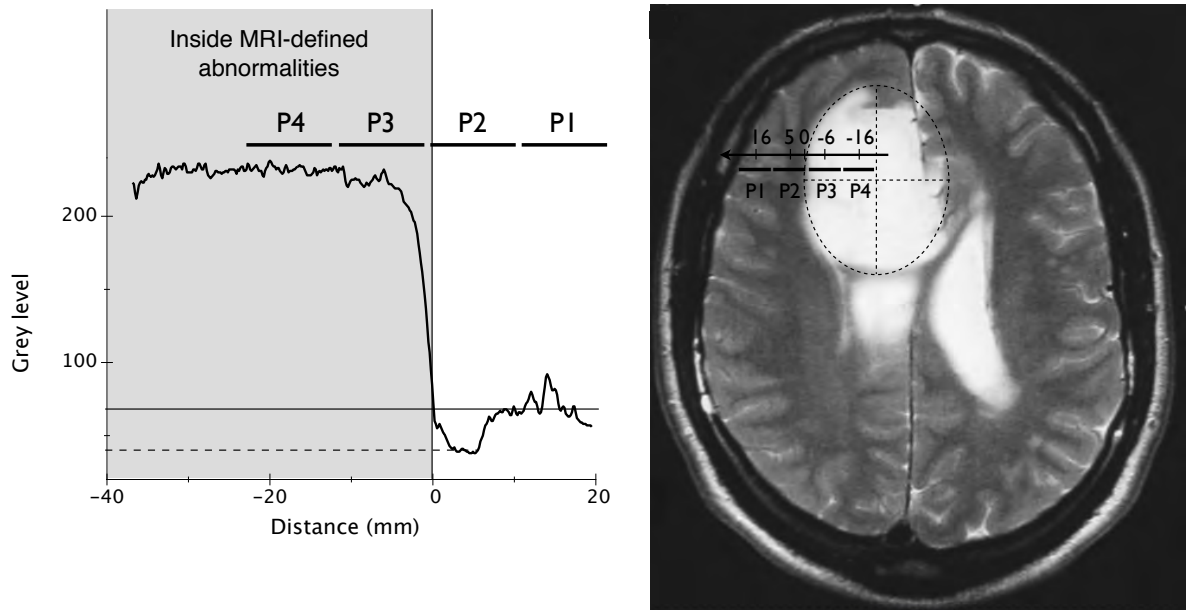

**Figure 1**: Right: MRI scan of patient 1, showing the location of the different biopsy samples along the biopsy trajectory and their position from the limit of the T2 signal abnormality (the samples beyond the equatorial plane of the tumor have been removed). The position 0 corresponds to the limit of the T2 hypersignal. The biopsy sample position has a positive value equal to the distance to the limit if the biopsy sample is outside the T2 hypersignal and a negative value, equal to the opposite of the distance to the limit, if the biopsy sample is inside. Left: Profile of the gray level along the biopsy trajectory. The limit of the abnormality has been defined as the position where the gray level in the tumor is equal to the mean value of the gray level in the cortex, see text and Figure 2 for more details. The gray rectangle indicates the region inside the limit of the MRI-defined abnormalities.

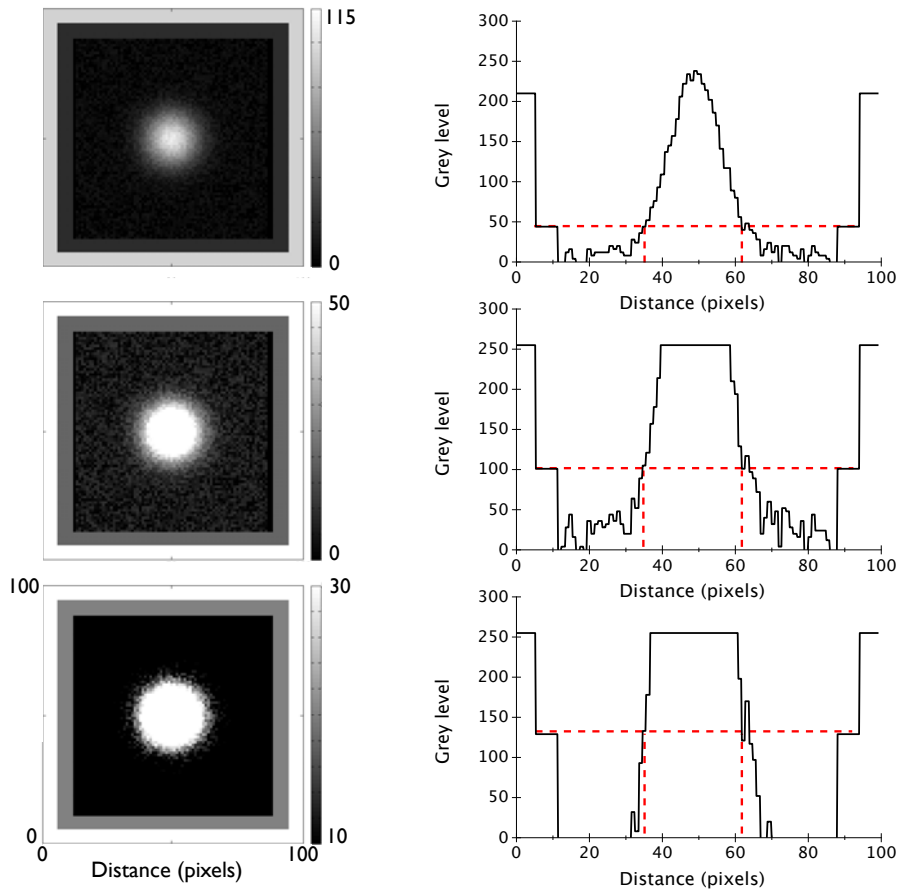

**Figure 2**: The size of the tumor on MRI scan does not depend on the contrast of the picture, if defined with a reference level (method of paragraph 3.2). We simulated a virtual T2-weighted MRI of a tumor-brain system (see text) discretized as a 100x100 matrix, then we chose three signal windows and plotted the corresponding MRI (left column): top, signal window [0, 115] (signal levels 0 or below are pictured as black, signal levels 115 or above are pictured as white); middle: signal window [0, 50]; bottom: signal window [10, 30]. Next to each picture, a graph of the gray level across a middle section of the image (y = 50, $0 \leq x \leq 100$) is shown. The signal abnormality is defined as the region (containing the image center) where the gray level is equal to or larger than the gray level of the cortex. When the signal window changes, the contrast between white and gray matter also changes, but the measured diameter of the abnormality remains the same, $26.5 \pm 1$ pixels.

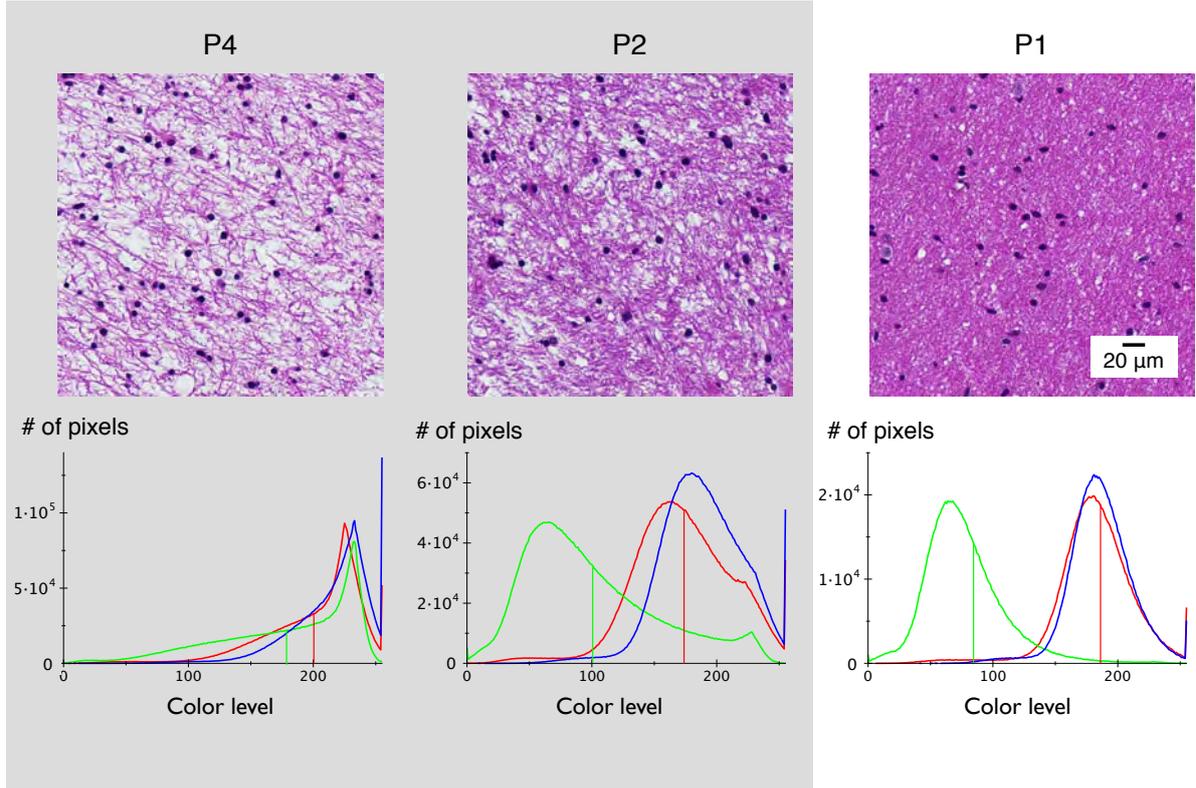

**Figure 3**: Color histograms (bottom) for tissues (top) with an increasing amount of edema from right to left. The gray rectangle indicates the region inside the limit of the MRI-defined abnormalities (samples P3 and P4). Right: this sample corresponds to P1, see figure 1, outside the MRI-defined abnormalities (the tissue is considered as normal). For the whole sample, the mean of the green histogram is $\overline{G_e} = 81$ and the mean of the red histogram is $\overline{R_e} = 182$. Middle: this sample corresponds to P3, inside the MRI abnormality limit, see Figure 1. For the whole sample, the mean of the green histogram is $\overline{G_e} = 101$ and the mean of the red histogram is $\overline{R_e} = 173$. Left: this sample corresponds to P4, inside the MRI abnormality limit (Figure 1). The mean of the green histogram is $\overline{G_e} = 177$ and the mean of the red histogram is $\overline{R_e} = 200$. The amount of edema can be calculated from the difference $\overline{R_e} - \overline{G_e}$ (see text). Analyzing all samples, we find that for this patient, the amount of edema is given by the formula: $x = 0.92(1.01 - 10^{-2}(\overline{R_e} - \overline{G_e}))$. For this patient, we obtain 0% of edema for sample P1, 26% for P3 and 71% for P4.

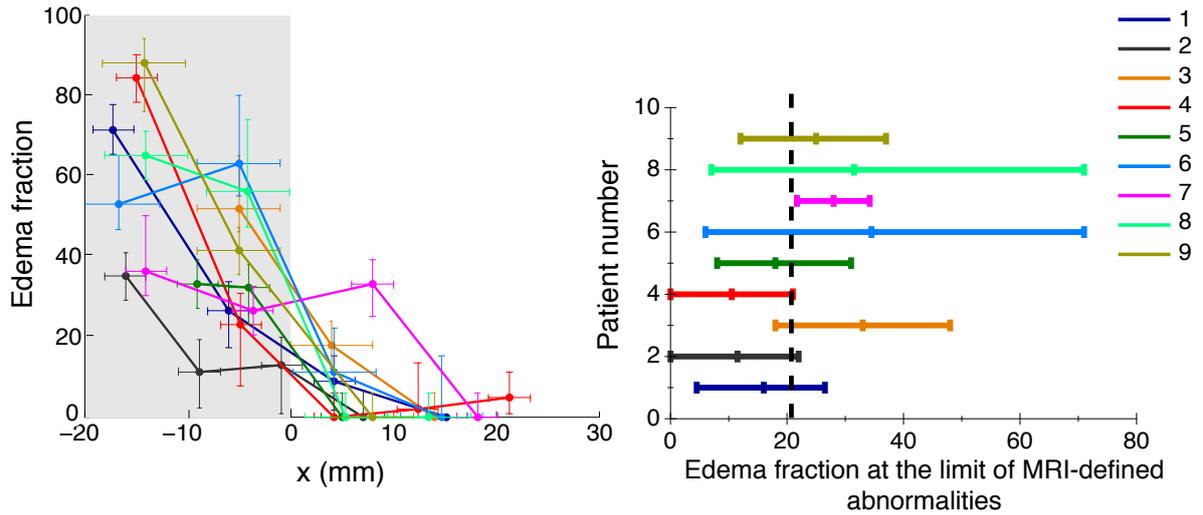

**Figure 4**: Left: edema fraction for all 9 patients, versus the position of the sample along the biopsy trajectory (the origin of the axis has been set at the limit of the MRI abnormality). The gray rectangle indicates the region inside the limit of the MRI-defined abnormalities. Right: edema fraction at the limit of the MRI abnormality for all patients.

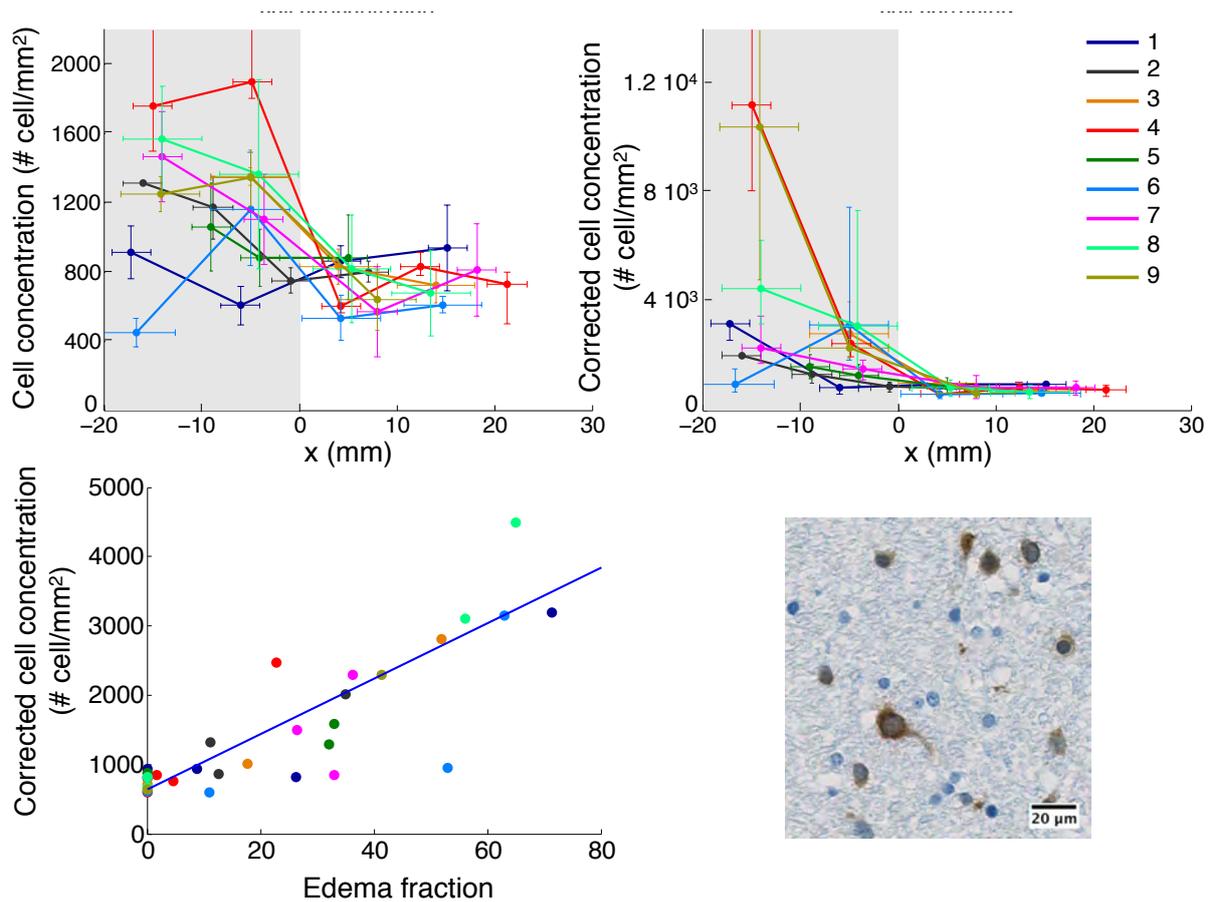

**Figure 5**: Top, left: cell concentration versus the position of the sample along the biopsy trajectory (the origin of the axis has been set at the limit of the MRI abnormality) for all patients. Top, right: corrected cell concentration versus the position of the sample along the biopsy trajectory: the concentration has been calculated with respect to the surface area of tissue (i.e. the total surface area of the sample minus the surface area occupied by edema), for all patients. The gray rectangles indicate regions inside the limit of the MRI-defined abnormalities. Bottom, left: Correlation between the corrected cell concentration and the edema fraction. The equation of the linear regression is y = 40.0 x + 635.2 and the determination coefficient is $R^2 = 0.78$. Bottom, right: A BS of patient 2 (situated at x = −16 mm), with IDH1 immunostaining, in order to measure the maximum section area of cells. The cytoplasm of IDH1 positive cells appears brown.

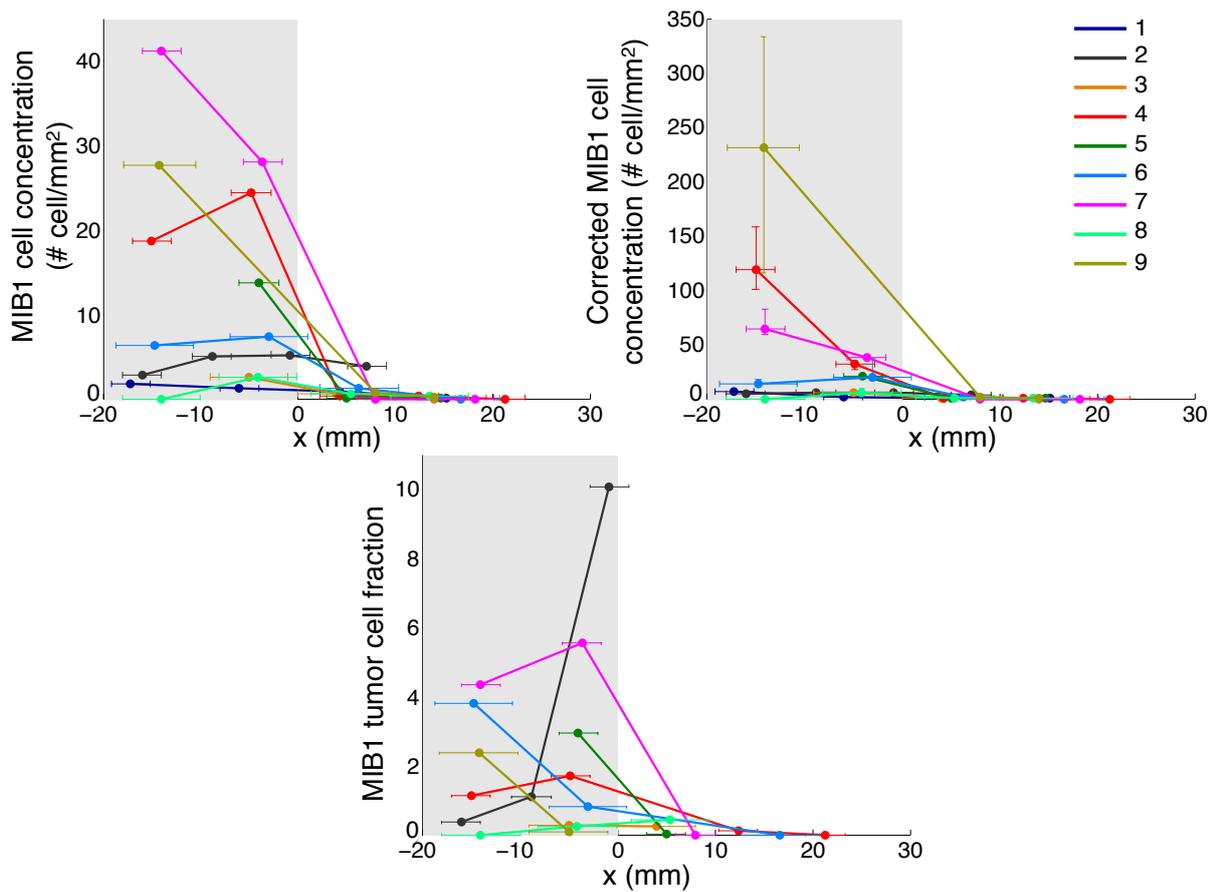

**Figure 6**: Top, left: MIB-1-positive cell concentration versus the position of the sample along the biopsy trajectory (the origin of the axis has been set at the limit of the MRI abnormality) for all patients. Top, right: corrected MIB-1-positive cell concentration versus the position of the sample along the biopsy trajectory: the concentration has been calculated with respect to the surface area of tissue (i.e. the total surface area of the sample minus the surface area occupied by edema). Bottom: ratio of the MIB-1-positive cell concentration to the difference between the cell concentration and the cell concentration outside the tumor. This quantity represents the fraction of tumor cells that are in proliferation. The gray rectangle indicates the region inside the limit of the MRI-defined abnormalities.

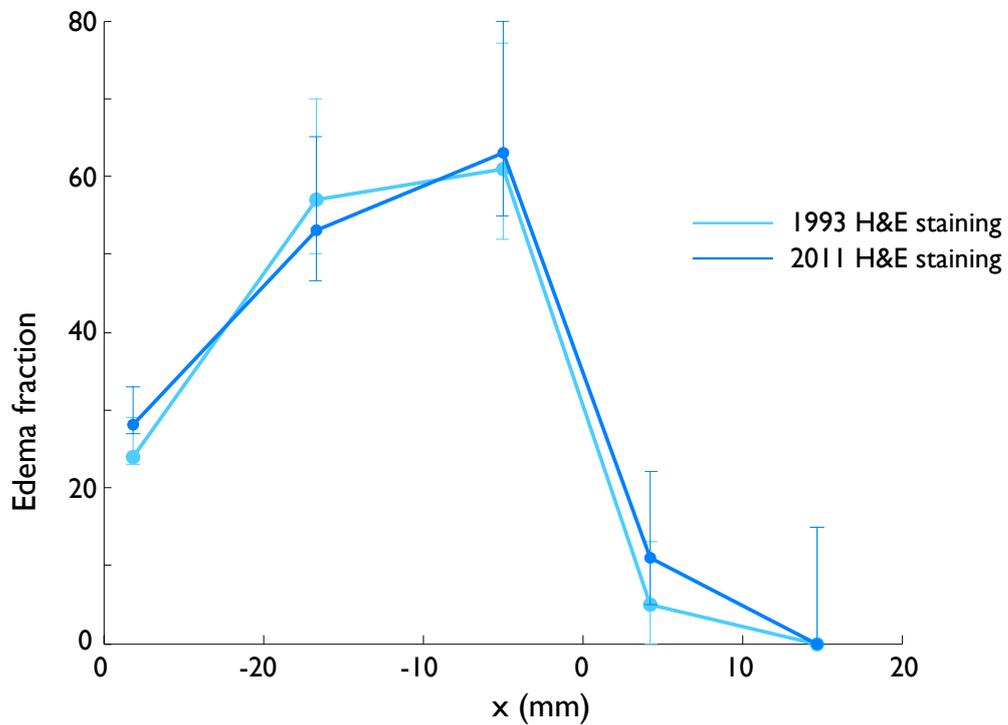

**Supplementary figure:**

For patient 6, comparison of edema fractions from old samples cut and stained in 1993 and new samples prepared in 2011. The dark blue dots and line correspond to 2011 samples and the light blue dots and line to 1993 samples. All the samples available for this patient appear on this figure (this is not the case on figure 4, 5, 6 where the inner samples, beyond the equatorial plane of the tumor, have been removed).